# Non-monotonic behaviour of the superconducting order parameter in Nb/PdNi bilayers observed through point contact spectroscopy

**P Romano**[1,2], **A Polcari**[1,2], **C Cirillo**[2,3] and **C Attanasio**[2,3]

[1]Dipartimento di Scienze per la Biologia, la Geologia e l'Ambiente, Università del Sannio, via Port'Arsa 11, 82100 Benevento, Italy.
[2]CNR-SPIN Salerno, via Ponte Don Melillo, 84084 Fisciano (SA), Italy.
[3]Dipartimento di Fisica "E.R. Caianiello" Università di Salerno, via Ponte Don Melillo, 84084 Fisciano (SA), Italy.

E-mail: promano@unisannio.it

**Abstract**
Point contact spectroscopy measurements have been performed on Nb/PdNi bilayers in which the thickness of the Nb layer, $d_{Nb}$, was kept constant to 40 nm while the thickness of PdNi, $d_{PdNi}$, was changed from 2 nm to 9 nm. Features related to the superconducting gap induced in the ferromagnet have been observed in the dV/dI versus V curves. These structures show a non-monotonic behaviour as a function of $d_{PdNi}$ as a consequence of the damped oscillatory behaviour of the superconducting order parameter in the ferromagnetic layer.

**PACS**. 74.45.+c; 74.78.Fk; 75.70.Cn.

## 1. Introduction

Recently great attention has been devoted to the coupling between superconductivity and ferromagnetism in Superconducting/Ferromagnetic (S/F) thin film hybrids [1,2] due to the rich physics originated from the coexistence of two competing orderings: in S the electrons with antiparallel spin are coupled to form Cooper pairs, while in F the exchange field, $E_{ex}$, forces the spin in a parallel configuration. For this reason a strong reduction of the order parameter in S/F hybrids is expected, since $E_{ex}$ will try to align the spins in the Cooper pair, leading to a strong pair breaking effect. Indeed, the proximity effect picture at the S/F interface is strongly modified compared to the S/N case (here N stands for normal metal). It is well known that the mechanism responsible of the proximity effect phenomenon at S/N interface is the Andreev reflection [3]. What happens is that electrons from the N side with energy lower than the superconducting gap, $\Delta$, cannot penetrate into the S side. However, an incoming electron can be transferred into the superconductor if a second electron is also transferred through the interface, thus forming a Cooper pair in S and creating a hole in N [3]. Electrons and holes will move in opposite directions, adding to the conductance of the normal electrode, but they will loose their coherence during the propagation in the metal. The Andreev pair disappears over a characteristic length $\xi_N$, which measures how far the two electrons leaking from the superconductor will diffuse in phase. In the dirty limit $\xi_N = (\hbar D_N/2\pi E)^{1/2}$, where $D_N$ is the diffusion coefficient of the normal metal and E is the energy responsible of the de-phasing. In a normal metal E is the thermal energy, $k_BT$, leading to $\xi_N = (\hbar D_N/2\pi k_BT)^{1/2}$. Andreev reflection thus contributes to an increasing of the conductance in the under-gap region, which will be well recognized in the dI/dV vs V curves. For example, in case of resistive contact, the conductance below the gap voltage ($\Delta$/e) becomes twice the normal conductance due to Andreev reflections. In the Blonder, Tinkham and Klapwijk (BTK) model [4], the conductance curves will show a characteristic bell-shaped behaviour with the gap value corresponding to the voltage where the conductance increases, at zero temperature, up to a maximum value of twice the background value. In other words, the differential resistance dV/dI of a pure resistive contact below the gap voltage becomes half of the normal resistance. In the model, an insulating (I) barrier mimics the properties of the interface between S and N, which strength can be progressively increased going from zero (pure resistive contact) up to reach a value that allows an appreciable tunneling current between S and N. As the barrier strength is increased the conductance will indeed show a double-peak structure at $\pm\Delta$/e, as in tunnel junctions, or double-dips in the resistance. In the case of an S/F boundary the situation is complicated due to energy splitting of the spin-up and spin-down sub-bands in the ferromagnet, which is responsible of a strong reduction of the reflections, since not all the majority spin electrons at the S/F boundary will be able to find a matching electron with opposite minority spin. For this reason the Andreev reflections are totally suppressed in fully spin polarized metal [5]. In addition, in S/F systems the coherence length, $\xi_F$, is strongly reduced. In this case, in fact, the incoming electron and the Andreev reflected hole occupy opposite spin bands. Consequently the induced superconducting order parameter disappears in F over a much shorter distance, this time controlled by the strength of the ferromagnet, namely, in the dirty limit, $\xi_F = (\hbar D_F/E_{ex})^{1/2}$, where $D_F$ is the diffusion coefficient of the ferromagnet. Moreover, due to the presence of the exchange field a spatial oscillation of the order parameter is superimposed on its decay in F [6].

The aim of this work is to probe this inhomogeneous character of the superconducting order parameter in S/F hybrids through Point Contact Spectroscopy (PCS). This technique has been widely used in the past to study the S electronic properties: mechanically pushing a tip, generally made of normal metal N, on the top of a superconductor, S-N contacts can be realized [7]. It is important to remind that the contact between tip and sample can also generate S-I-N and S-I-S contacts, when a suitable barrier, either natural or artificial, is present between the two electrodes [8-11]. From the I-V behaviour, and more specifically from the dI/dV curves, it is possible to estimate the basic properties of a superconductor, as the energy gap value and the density of states, as well as the symmetry of the order parameter in the case of unconventional superconductors, being the experiment sensitive to both the magnitude and the phase of the order parameter [7]. While for the high-$T_c$ cuprates, PCS provided the earliest measurements of the superconducting gap spectra [8], more recently it also appeared to be a powerful tool in investigating the superconducting order parameter even in the case of multiple gaps [12]. Moreover, using a superconducting tip, S-F contacts have been successfully employed to determine the spin polarization of several ferromagnets [13]. PCS technique has also be extended to study proximized structures, like S/F bilayers, obtaining S/F-N contacts between the bilayer and the normal metal tip [14]. Since in this case two different interfaces are involved, namely between S and F, and between F and N, more complex conductance curves may be obtained. The BTK model should then include these processes, as well as the presence of polarized electrons due to F. For all these reasons, while PCS can be simple in ideal situation, in real experimental conditions its application, as well as the data interpretation, can be quite complicated.

In this paper we study the simplest S/F hybrids, namely S/F bilayers. For these structures a non-monotonic behaviour of the critical temperature, $T_c$, over $\xi_F$ as a function of the thickness of the ferromagnetic layer, $d_F$, has been found theoretically as well as experimentally [1,15-20]. The superconducting transition temperature is the simplest parameter which reveals intriguing behaviour typical of S/F structures, but the inhomogeneous character of the order parameter in the ferromagnetic layer also affects the characteristics of S/F/S Josephson junctions [21], as well as the density of states (DOS) in S/F based tunnel junctions [22,23]. Following this approach, we present PCS measurements realized between a normal metal Au tip on Nb/PdNi bilayers with different PdNi thickness, PdNi being a weakly ferromagnetic alloy. We observe that some features are present in the dV/dI versus V curves which are related to the superconducting gap induced in the ferromagnet as well as to the Nb order parameter. Both these structures show a non-monotonic behaviour as a function of $d_{PdNi}$.

## 2. Fabrication

The S/F bilayers consist of a 40-nm thick Nb layer, and of a weakly ferromagnetic alloy, $Pd_{0.84}Ni_{0.16}$ (=PdNi) layer, with variable thickness, $d_{PdNi}$ = 2-4-5-9 nm. This system was chosen since the existence of an inhomogeneous superconducting order parameter in F, peculiar characteristic of S/F hybrids [1,2,6,15,16], has been already demonstrated for Nb/$Pd_{1-x}Ni_x$ systems [19-22,24-26]. Moreover, PdNi is characterized by longer spin-flip scattering length compared to the widely used CuNi alloy [27]. The bilayers have been deposited by a three-target UHV dc magnetron sputtering, equipped with a load-lock chamber. The base pressure in the main chamber was in the low $10^{-8}$ mbar range. The Nb and PdNi layers have been grown on $Al_2O_3$ substrates at typical power of $W_{Nb}$=390 Watt and $W_{PdNi}$=90 Watt, and Argon pressure of 3 μbar and 8 μbar, for Nb and PdNi respectively. During the deposition the substrate holder was kept at T=100 °C. These fabrication conditions determine the deposition rates $r_{Nb}$ = 2.7 nm/s and $r_{PdNi}$ = 2.2 nm/s, respectively, which were controlled with a thickness monitor calibrated by low angle X-Ray reflectivity measurements. The Ni content of the ferromagnetic alloy, x=0.16, has been determined by Energy Dispersion Spectroscopy measurements. As reported in detail in Ref. [25] for this composition, the ordering temperature and the exchange energy of the alloy are $T_{Curie}$=190 K and $E_{ex}\approx$14 meV, respectively. Using for the diffusion coefficient in the ferromagnet $D_F$ = 2.3 x $10^{-4}$ $m^2$/s [25] we have a penetration of the Cooper pairs inside the F layer $\xi_F$ of the order of 3-4 nm.

## 3. Preliminary characterization

The bilayers have been preliminary characterized by transport measurements. In particular the superconducting transition temperature was resistively measured using a standard dc four-probe technique. $T_c$ was defined at the midpoint of the R(T) transition. The transitions width, defined as $\Delta T_c = T_c(0.9R_N) - T_c(0.1R_N)$, where $R_N$ is the value of the resistance in the normal state, never exceeded 50 mK. The single Nb film, 40 nm thick, has a superconducting temperature $T_{cS} = 8.2$ K, a low temperature resistivity $\rho_{Nb}=12$ $\mu\Omega$cm, and a superconducting coherence length, estimated from perpendicular upper magnetic field, $\xi_S = 6$ nm. The dependence of $T_c$ as a function of the PdNi layer thickness in Nb/PdNi bilayers is reported in figure 1. The critical temperature rapidly decreases with increasing the thickness of the ferromagnetic layer, showing a minimum around $d_{PdNi} \approx 5$ nm. The value of the dip position scales reasonably well with the ones obtained for Nb/PdNi systems for different Ni concentrations of the alloy [19,20,24]. Despite the small number of samples, this result confirms the non-monotonic behaviour of $T_c(d_{PdNi})$, the difference in the critical temperature between the sample with $d_{PdNi} = 5$ nm and the saturation value being in fact $\Delta T \approx 0.14$ K >> $\Delta T_c$. As discussed in the following section, the signatures of the non-homogeneous superconducting order parameter were further investigated below $T_c$ by PCS technique.

## 4. Point contact measurements

In order to infer the electronic properties in our samples, we have performed point contact measurements making possible the process of Andreev reflections between a normal tip and the bilayer. A tip made of normal metal (Au) has been pushed on the ferromagnetic side of a S/F bilayer. Since Au is a soft material, a direct contact with Nb can be excluded, the tip flattening rather than damaging the PdNi layer. To shed light on the effects of the tip-sample interaction Scanning Electron Microscopy (SEM) analyses have been performed. The results (not reported here) reveal in fact no sign of degradation and/or presence of holes in the PdNi surface. The I-V curves of the contact were recorded while the junction was current biased and the gap value was obtained from the numerically calculated resistance dV/dI for each sample. A sketch of the spatial dependence of the order parameter, $\Psi(x)$, in S/F bilayers for two different values of $d_F$ is reported in figure 2, where also the point contact geometry is illustrated. The Andreev spectroscopy measurements should be performed in the ballistic limit, namely for values of the point contacts size much smaller than the carrier mean free path. The point contact diameter, $a$, can be evaluated by means of the Sharvin resistance in the ballistic regime as [28]

$$R = 4\rho l / 3\pi a^2 \qquad (1)$$

where $\rho$ and $l$ are the low temperature resistivity and mean free path of the sample. In our case the measured value of $\rho$ was equal to 13 μΩcm for all the investigated bilayers. Since this value is very close to the measured low temperature resistivity for a single Nb film ($\rho_{Nb}$ = 12 μΩcm) we assume that $\rho_{Nb/PdNi} l_{Nb/PdNi}$ = 3.72x10$^{-6}$ μΩcm$^2$, value which is valid for single Nb [29]. Within this limit we obtain $l$ = 3 nm. The resistance of the contacts in our samples, of the order of few Ohms, gives from Eq. (1) a diameter ranging from 10 to about 100 nm, larger than the mean free path of carriers in the bilayer. Therefore, this large contact area can cause many other effects as diffusive transport, multiple contacts and proximity effects [30], which are not intrinsic features of the samples and can affect the spectra. On the other hand, the Sharvin formalism rigorously applies for clean systems [31], while in dirty superconductors with extremely low $l$ values the condition $a<l$ is hardly achieved. When the contact resistance is measured, this could just reflect in a higher effective barrier strength but still in the framework of a ballistic transport [31].

In figure 3 the differential resistances numerically calculated from the I-V curves of three samples with different ferromagnetic thickness, $d_F$, are shown below $T_c$. The corresponding reduced temperature, $t=T/T_c$, is indicated for each curve. The general behaviour is characterized by two components at low bias: a sharp single dip at zero and double symmetric dips indicated in figure by the red arrows. The zero-bias dip, known as zero-bias anomaly (ZBA), has also been observed in different superconducting systems such as Nb-Ag (or Al) microjunctions [32] and microshort Au-$YBa_2Cu_3O_{7-\delta}$ junctions [33]. Most notably, ZBA seems to appear more systematically in the Andreev spectra with small contact resistances and thus with large contact sizes, indicating that it can be related to the contact geometry and not an intrinsic feature of the sample in question. Various theoretical models have been proposed, like ordinary proximity effects [33] or multiple Andreev reflections [34], both predicting easy suppression of the effect by a small applied magnetic field. In our samples, the ZBA decreases with a magnetic field applied perpendicularly to the sample, as shown for example in figure 4a for a contact realized on the sample with $d_F$ = 9 nm at T=2.5 K. The zero-bias conductance normalized to the background conductance, $G(0)_{norm}$, versus the applied magnetic field is shown in figure 4b:

$G(0)_{norm}$ rapidly decreases down to 0.04 T, a value comparable with the coercive field of PdNi [25], and totally disappears already at 0.1 T. The rapid reduction at low field indicates that Nb is not responsible for this behaviour, being the typical upper critical magnetic field values at this temperature for our samples of the order of 1 T [24]. On the contrary, it seems reasonable to ascribe this dependence to the occurrence of crossed Andreev reflections from multiple F domains close to the S/F interface [35]. On the other hand, as shown in figure 5 for two different contacts realized on the same sample, the ZBA disappears in zero-field when the temperature is increased up to 4.2 K, value which is smaller than the critical temperature of the sample. This behaviour is probably due to the local nature of the PCS technique, which is only able to probe a surface portion of the sample. It is worth noting that, regardless its origin, in some cases the ZBA can overwhelm the entire conductance spectrum and render the gap measurements unfeasible [36]. In our case, the ZBA might be related to a change of phase of the order parameter in F. When, in fact, the order parameter changes sign along the electron's trajectory, a zero-bias conductance peak can appear due to Andreev surface bound states (ABS). This feature, absent in BTK's calculation, is for instance expected for the case of a gap having a d-wave symmetry, as Kashikawa et al. have demonstrated for YBCO [37]. Even more relevant for our work, the presence of ABS is expected also for S/F systems [38].

Turning our attention to the dV/dI data, the dips just above the ZBA have been related to the superconducting gap voltage in S-N microjunctions [32], as expected in the BTK model. In our case, the double dip structure could be the signature of the superconducting gap induced in the F layer, for this reason much lower than the Nb gap value. In figure 6 the double voltage dip position $V_{dip}$ evaluated from figure 3 is shown as a function of $d_F$. The value for $d_F=0$ has been calculated from the $T_c$ of pure Nb through the relation $2\Delta/k_BT_c=3.52$. Although for the sample with $d_F = 2$ nm the value of reduced temperature is different (t=0.56), the dip position can still be reasonably related to the zero-temperature gap voltage, considering the temperature dependence of the superconducting gap in conventional materials [39]. A non-monotonic behaviour of $2V_{dip}(d_F)$ is observed, very close reminding the $T_c(d_F)$ dependence of figure1 (reported again in figure 6 for sake of clarity), with a clear minimum at $d_F = 5$ nm which represents the experimental confirmation of the change of sign of the superconducting order parameter at the F/I interface (I is the vacuum) [15, 38]. The additional resistance dips sometimes observed in the low bias region at voltages $V_n$, probably arise from n-fold Andreev reflection process [39], or from multiple scattering and impurities [40].

Finally, in the dV/dI data presented in figure 3 one more feature can be observed, namely a resistance double peak in the high bias region which appears stable and symmetric around zero (black arrows in the figure) in all the samples. Similar features have been observed on S-N microcontacts [32] as well as in heterojunctions involving contacts between a superconductor and a semiconductor [41]. Although their origin is not completely clear, these structures, appearing like minima in the conductance of Andreev contacts, have been explained in terms of proximity-induced superconductivity [42]. In figure 7 we report the double voltage peaks position, $V_{peak}$, as a function of $d_F$, which shows, once again, a non-monotonic behaviour, even though in this case, probably due to the few experimental points, the minimum of the curve appears at $d_F$ = 2 nm. We can also estimate the critical temperature dependence, $T_c^{2Vpeak}(d_F)$, from the position of the dip through the relation $2V_{peak}/k_B T_c^{2Vpeak} = 3.52$. As it is shown in the inset of figure 7 the values for $T_c^{2Vpeak}$ are slightly different from the values of the critical temperature measured from transport measurements. We believe that this discrepancy is again due to the fact that PCS is a local measurement while resistive transitions probe the average superconducting order parameter over the whole sample.

## 5. Discussion and Conclusions

Nb/PdNi bilayers for different values of PdNi thickness have been analyzed. The $T_c$ of the bilayers rapidly decreases with increasing the PdNi thickness, showing a minimum for $d_{PdNi}$ of about 5 nm. The non-monotonicity exhibits itself also in PCS measurements. In fact, a non-monotonic behaviour of the gap induced in the ferromagnet, connected with a resistance double dip structure, is observed. A similar result is found for the proximized Nb order parameter, connected with a resistance double peak structure, as a function of $d_{PdNi}$. The conventional BTK theory [4] cannot help as it is to model the data, since it analyzes the conductance versus bias voltage curves for nonmagnetic metal-superconductor contacts. In our samples, the superconductor has been substituted by a S/F bilayer, making the situation somewhat different. In this case, in fact, proximity can occur between S and F. As a consequence, new features can appear in the data, which can be accounted only through a modification of the model, for instance including two gaps in the conductance expression, one for the Andreev process and one for the quasiparticle transport [42]. Moreover, the spin polarization of the ferromagnet should also be considered in the model; as a further consequence, the normalized conductance at zero bias will be lower than 2, as instead expected for pure Andreev reflections, due to an imbalance in the number of spin-up and spin- down electrons which can cause a suppression of the Andreev reflection probability. In the data, we have also observed a ZBA in the form of a zero bias dip. This feature, absent in BTK's calculations, might be explained in terms of a change of sign of the order parameter along the electron's trajectory. For high-$T_c$ YBCO superconductor, for instance, it has been shown that a ZBA can come from an unconventional gap symmetry (d-wave) [37], due to the fact that the quasiparticles may experience a sign change of the order parameter. The general consequence is the development of ABS in the quasiparticle excitation spectrum at the Fermi energy. These bound states exist only at the interface but do not have important effect on the overall measured spectrum, leading to a ZBA in certain geometries. The BTK model can thus be extended in order to describe these novel effects. In the system under study the ZBA can originate from a change of sign of the order parameter at the F/I interface [38]. Despite the small number of the analyzed samples, the experimental data presented in this work confirm the change of sign of the superconducting order parameter at the S/F interface as a function of the ferromagnetic layer thickness. More work is in progress in order to further analyze our data by means of appropriate models, taking into account the new features appearing in the resistance curves and the role of the ferromagnet, in the framework of the interactions between superconductivity and ferromagnetism.

**Acknowledgements**

The authors wish to thank Dr. R. Fittipaldi for performing SEM analyses.

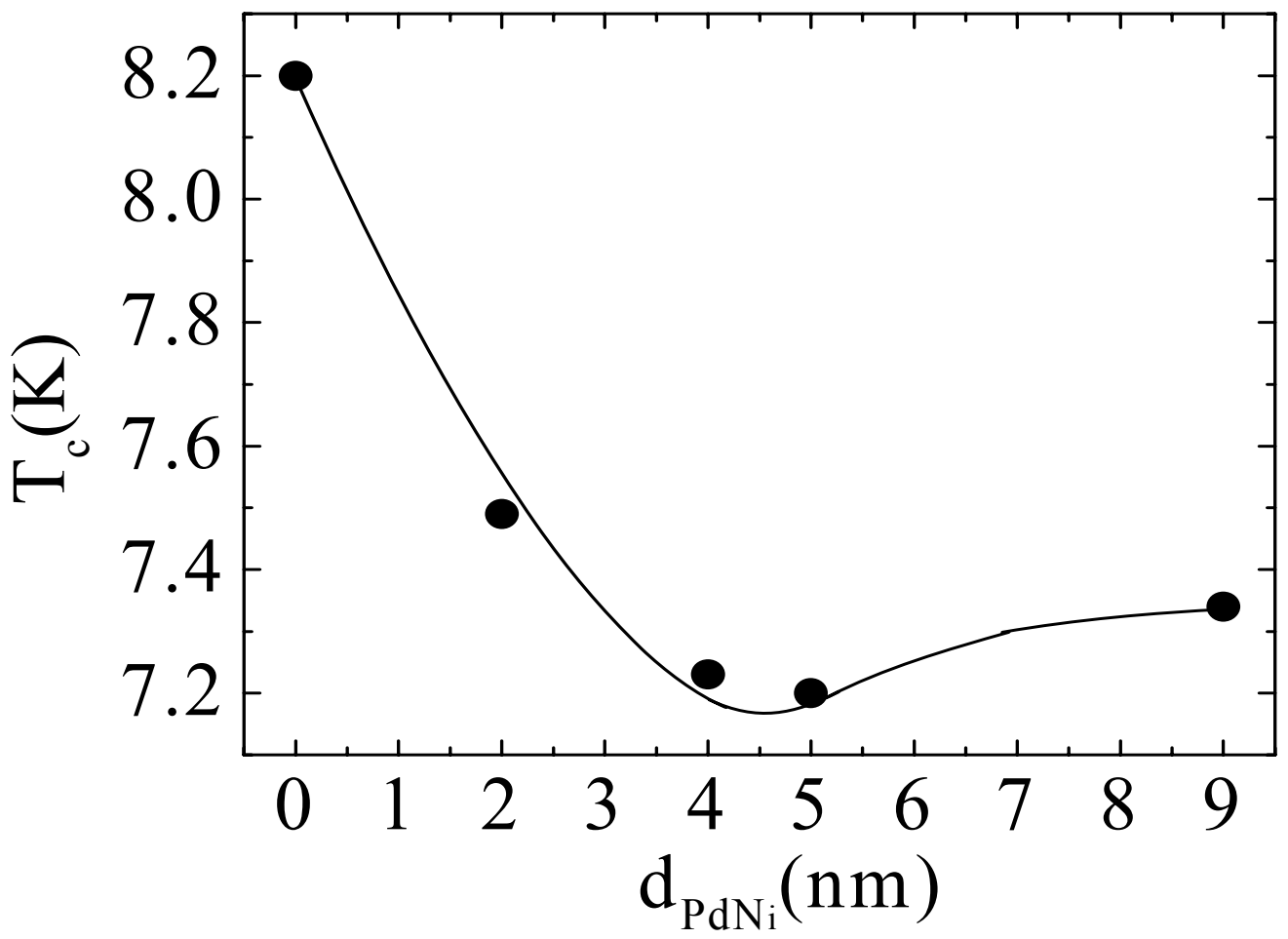

**Figure 1**. Superconducting critical temperature, $T_c$, as a function of the PdNi thickness, $d_{PdNi}$. The line is a guide to the eye.

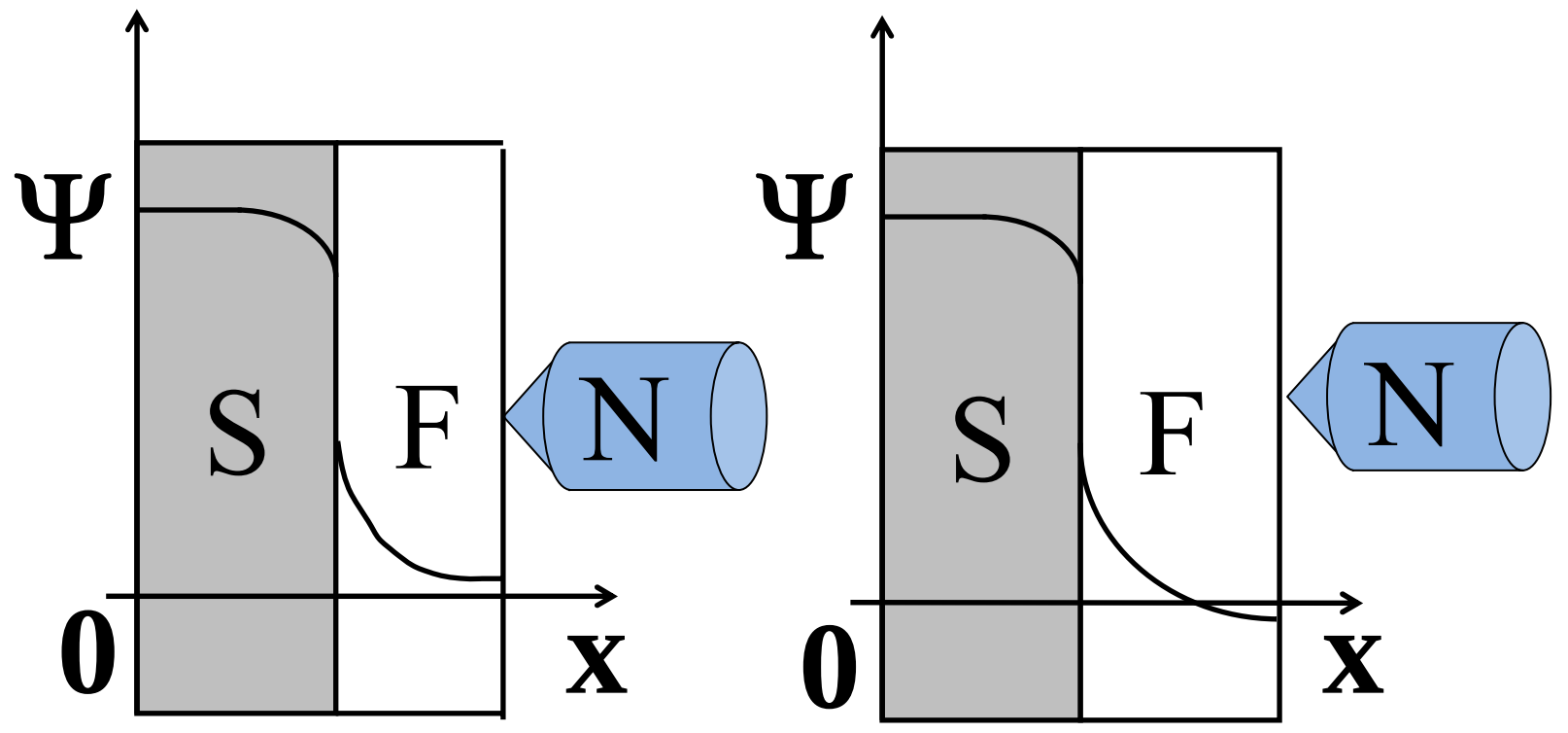

**Figure 2**. (Color online) Sketch of the spatial dependence of the order parameter, Ψ(x), in S/F bilayers for two different values of $d_F$ (adapted from Ref. [15]). Depending on the ferromagnetic layer thickness, the sign of the order parameter at the F free surface may change from positive to negative. A scheme of the point contact measurement geometry is also illustrated. The whole sketch is not in scale.

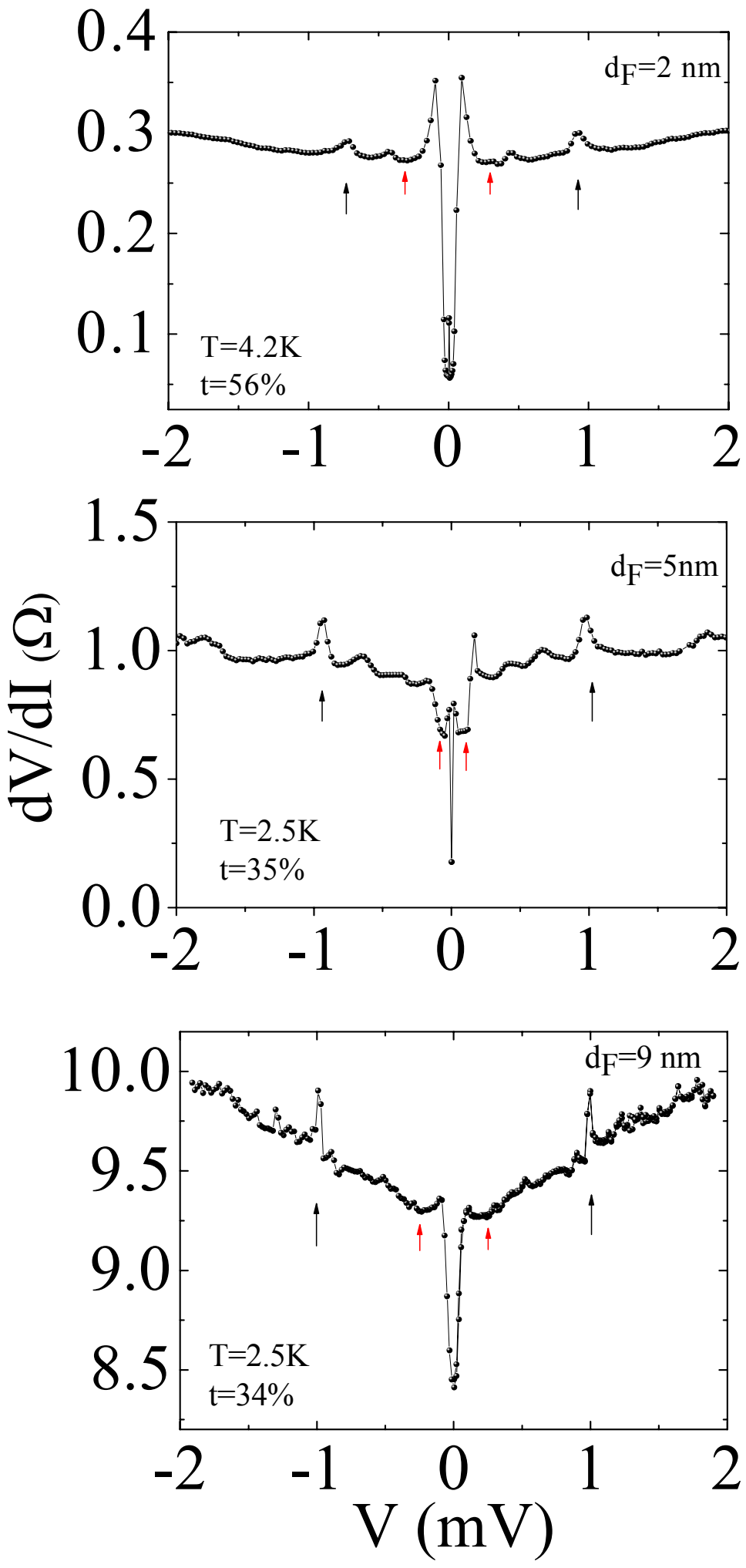

**Figure 3**. (Color online) Differential resistances, dV/dI, for samples with different thickness of the PdNi layer, $d_F$, at different reduced temperatures, $t=T/T_c$. The red arrows indicate a double symmetric dip just above the ZBA. The black arrows indicate a double symmetric peak at higher biases.

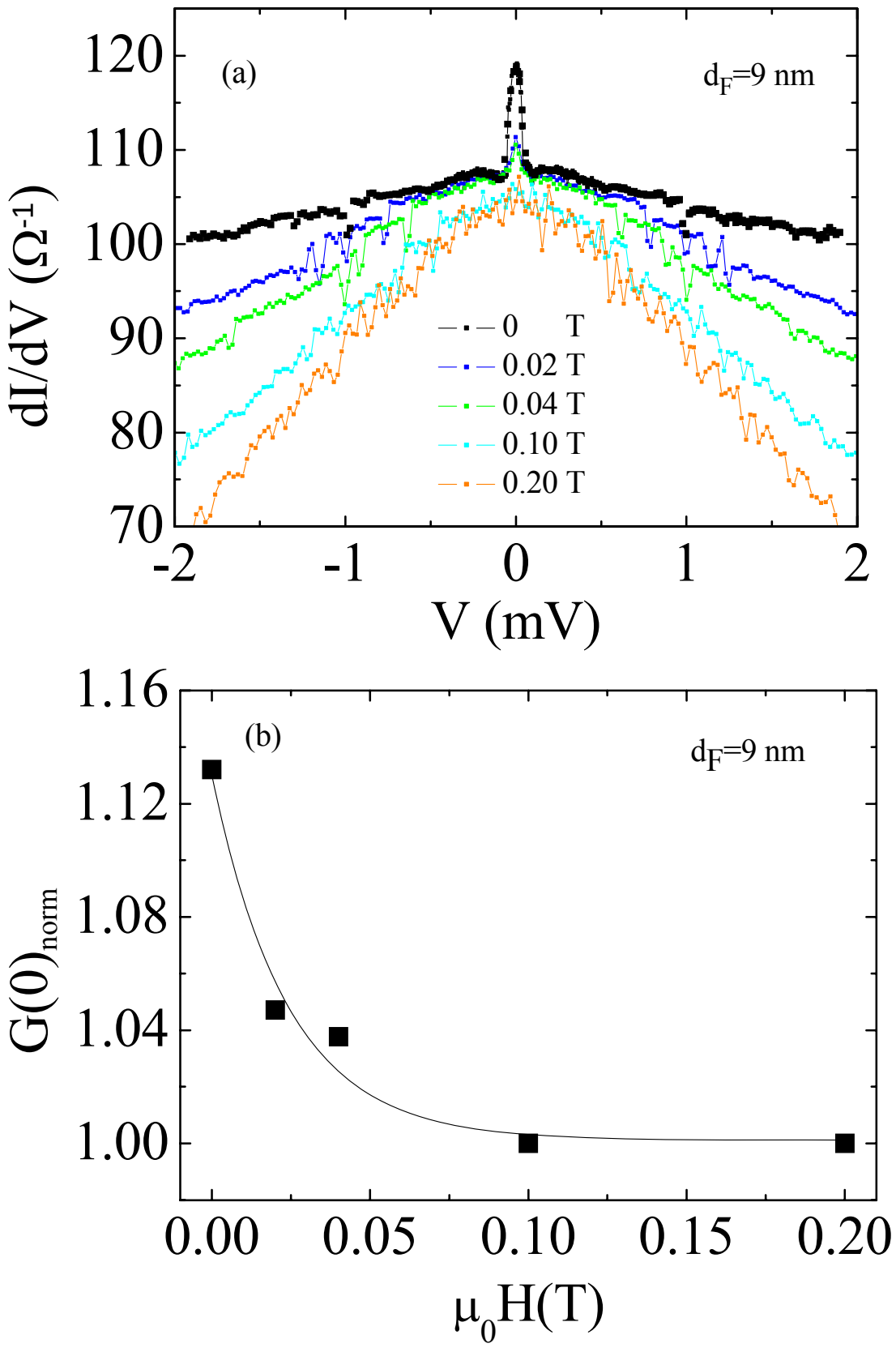

**Figure 4**. (Color online) The conductance of a contact realized on the sample with $d_F$ = 9 nm at different fields. b) The normalized zero-bias conductance as a function of the applied magnetic field.

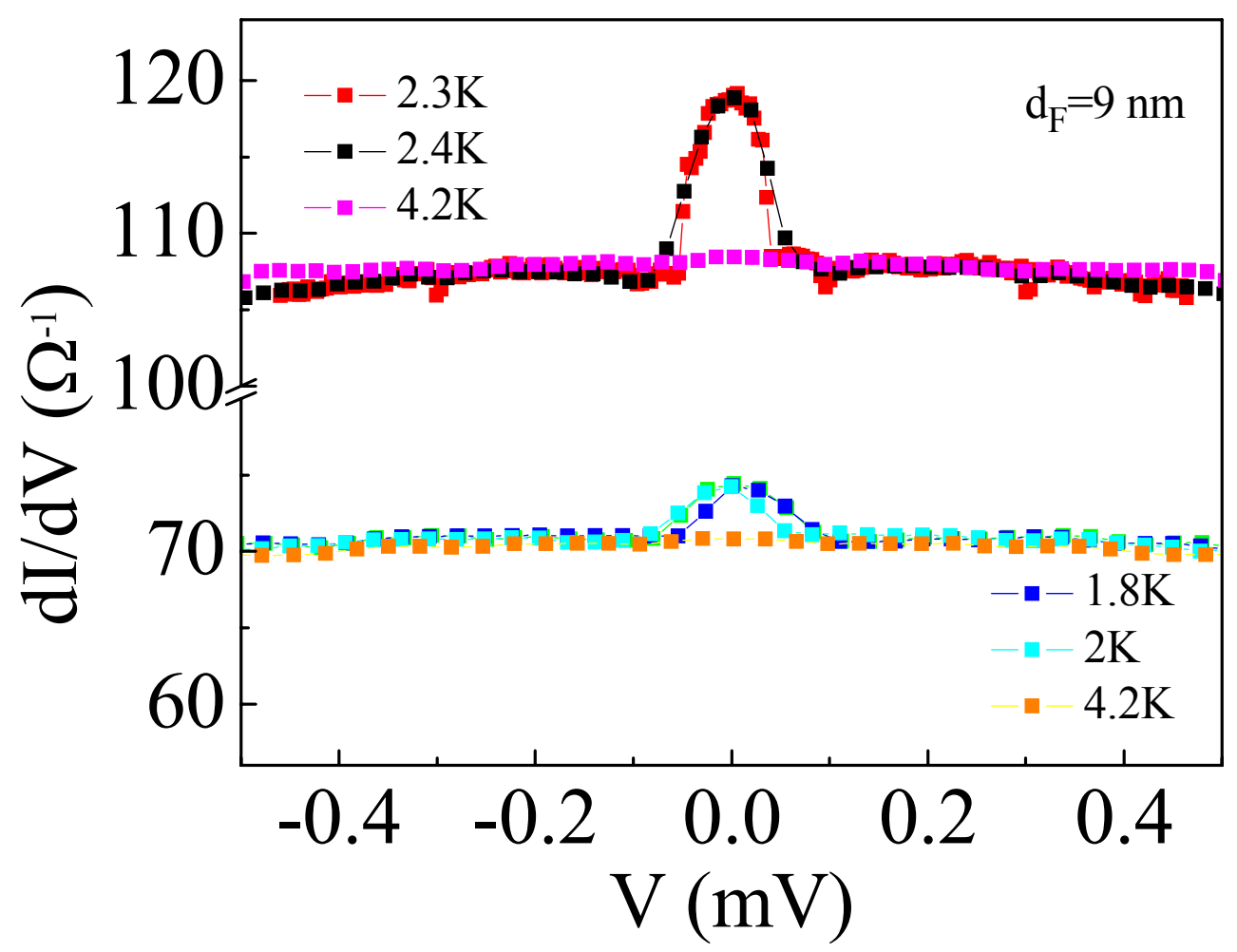

**Figure 5**. (Color online) The conductance of two different contacts realized on the sample with $d_F$ = 9 nm at different temperatures.

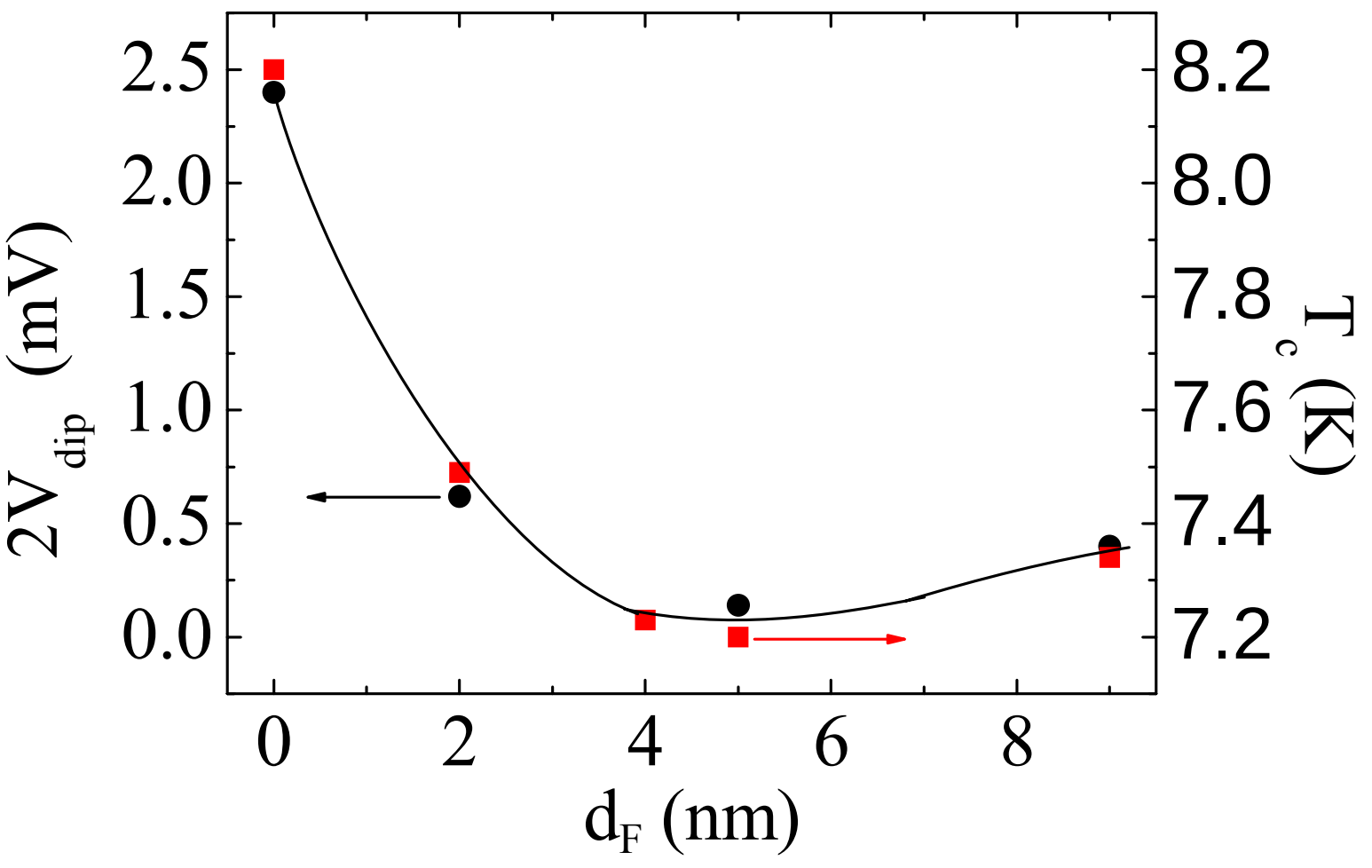

**Figure 6**. (Color online) Thickness dependence of the low bias resistance double dip position as indicated by the red arrows in figure 3 (left scale) and of the critical temperature values (right scale). The line is a guide to the eye.

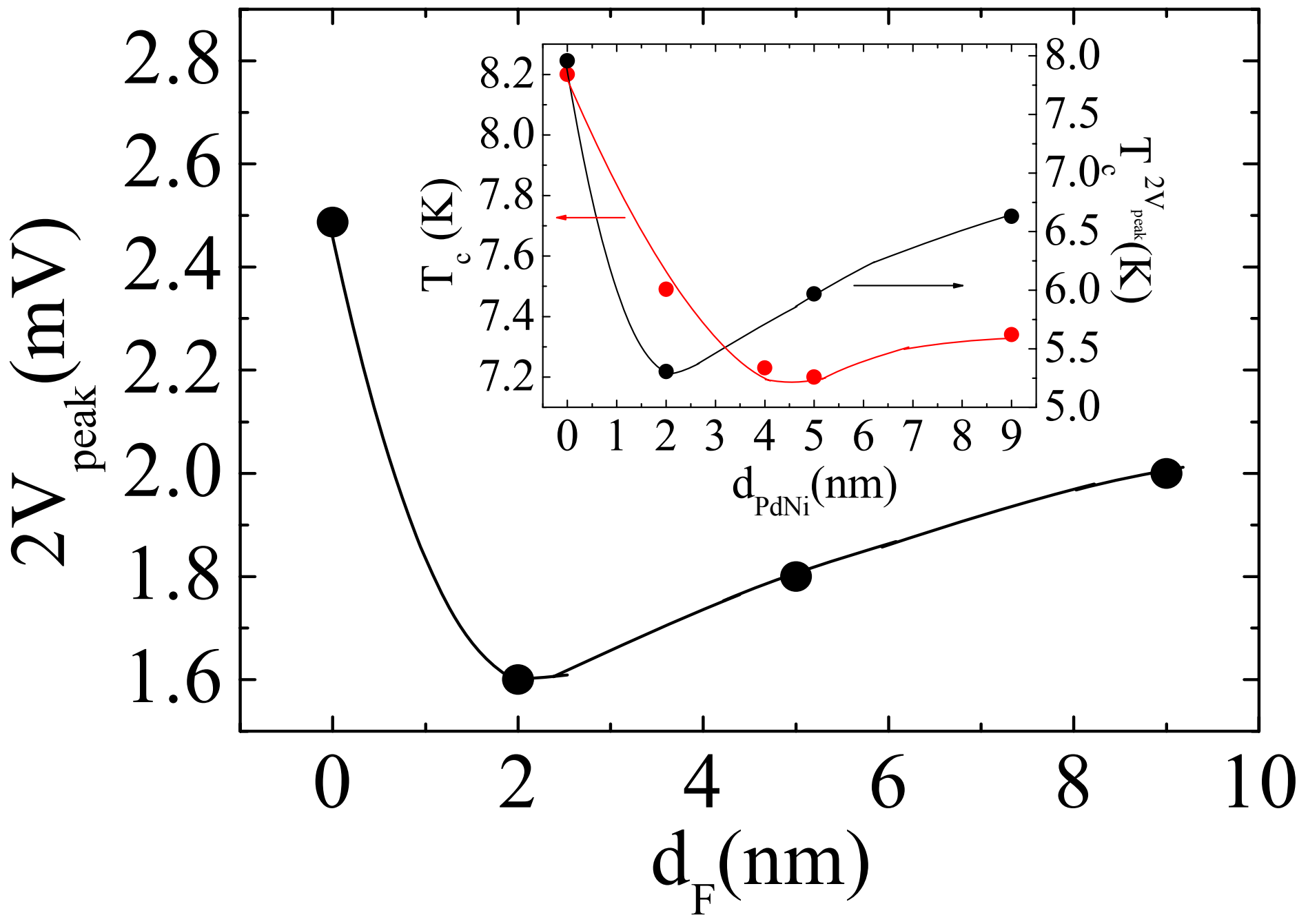

**Figure 7**. (Color online) High bias resistance double peak position as a function of $d_F$ (see black arrows in figure 3). Inset: comparison between the critical temperature values estimated from the high bias double peak position (right scale) and from the R(T) measurements (left scale). The lines are guide to the eye.

## References

[1] Buzdin A I 2005 *Rev. Mod. Phys.* **77** 935

[2] Bergeret F S, Volkov A F and Efetov K B 2005 *Rev. Mod. Phys.* **77** 1321

[3] Andreev A F 1964 *Zh. Eksp. Teor. Fiz.* **46** 1823 [1964 *Sov. Phys. JETP* **19** 1228]

[4] Blonder G E, Tinkham M and Klapwijk T M 1982 *Phys. Rev. B* **25** 4515

[5] de Jong M J M and Beenakker C W J 1995 *Phys. Rev. Lett.* **74** 1657

[6] Demler E A, Arnold G B and Beasley M R 1997 *Phys. Rev. B* **55** 15174

[7] Deutscher G 2005 *Rev. Mod. Phys.* **77** 109

[8] Zasadzinski J 2003 in The Physics of Superconductors, Ed. by Bennemann K H and Ketterson J B (Springer, Berlin)

[9] Huang Q, Zasadzinski J F , Gray K E, Richards D R and Hinks D G 1990 *Appl. Phys. Lett.* **57** 2356

[10] DeWilde Y, Miyakawa N, Guptasarma P, Iavarone M, Ozyuzer L, Romano P, Hinks D G, Kendziora C, Crabtree G W and Gray K E 1998 *Phys. Rev. Lett.* **80** 153

[11] Romano P, Chen J and Zasadzinski J 1998 *Physica C* **295** 15

[12] Szabó P, Samuely P, Kacmarcík J, Klein T, Marcus J, Fruchart D, Miraglia S, Marcenat C and Jansen A G M 2001 *Phys. Rev. Lett.* **87** 137005; Chen T Y, Tesanovic Z, Liu R H, Chen X H, and Chien C L 2008 *Nature* **453** 1224

[13] Soulen R J Jr, Byers J M, Osofsky M S, Nadgorny B, Ambrose T, Cheng S F, Broussard P R, Tanaka C T, Nowak J, Moodera J S, Barry A and Coey J M D 1998 *Science* **282** 85

[14] Piano S, Bobba F, De Santis A, Giubileo F, Scarfato A and Cucolo A M 2006 *Journal of Physics: Conference Series* **43** 1123

[15] Fominov Y V, Chtchelkatchev N M and Golubov A A 2002 *Phys. Rev. B* **66** 014507

[16] Tagirov L R1998 *Physica C* **307** 145

[17] Garifullin I A, Tikhonov D A, Garifyanov N N, Fattakhov M Z, Tagirov L R, Theis-Bröhl K, Westerholt K and Zabel H 2004 *Phys. Rev. B* **70** 54505 and references therein.

[18] Sidorenko A S, Zdravkov V I, Kehrle J, Morari R, Obermeier G, Gsell S, Schreck M, Müller C, Kupriyanov M Yu, Ryazanov V V, Horn S, Tagirov L R and Tidecks R 2009 *JETP Lett.* **90/2** 139 [2009 *Pis'ma v ZhETF* **90** 149] and references therein

[19] Cirillo C, Rusanov A, Bell C and Aarts J 2007 *Phys. Rev. B* **75** 174510

[20] Cirillo C, Prischepa S L, Salvato M, Attanasio C, Hesselberth M and Aarts J 2005 *Phys. Rev. B* **72** 144511

[21] Kontos T, Aprili M, Lesueur J, Genet F, Stephanidis B and Boursier R 2002 *Phys. Rev. Lett.* **89** 137007

[22] Kontos T, Aprili M, Lesueur J and Grison X 2001 *Phys. Rev. Lett.* **86**, 304

[23] Crétinon L, Gupta A K, Sellier H, Lefloch F, Fauré M, Buzdin A and Courtois H 2005 *Phys. Rev. B* **72**, 024511

[24] Cirillo C, Bell C, Iannone G, Prischepa S L, Aarts J and Attanasio C 2009 *Phys. Rev. B* **80** 094510

[25] Cirillo C, Ilyina E A and Attanasio C 2011 *Supercond. Sci. Technol.* **24** 024017

[26] Kushnir V N, Prischepa S L, Aarts J, Bell C, Cirillo C and Attanasio C 2011 *Eur. Phys. J. B* **80** 445

[27] Arham H Z, Khaire T S, Loloee R, Pratt W P Jr. and Birge N O 2009 *Phys. Rev. B* **80** 144515 and references therein

[28] Sharvin Yu V 1965 *Sov. Phys. JEPT* **21** 655 [1965 *Zh. Eksp. Teor. Fiz.* **48** 984]

[29] Minhaj M S M, Meepagala S, Chen J T and Wenger L E 1994 *Phys. Rev. B* **49** 15235

[30] Rajanikanth A, Kasai S, Ohshima N and Hono K 2010 *Appl. Phys. Lett.* **97** 022505

[31] Heslinga D R, Shafranjuk S E, van Kempen H and Klapwijk T M 1994 *Phys. Rev.* B **49** 10484

[32] Xiong P, Xiao G and Laibowitz R B 1993 *Phys. Rev. Lett.* **71** 1907

[33] Srikanth H and Raychaudhuri A K 1992 *Phys. Rev.* B **45** 383

[34] Volkov A F and Klapwijk T M 1992 *Phys. Lett. A* **168** 217

[35] Bakaul S R, Li K, Han G and Wu Y 2008 *IEEE Trans. On Magn.* **44** 2737

[36] Chen T Y 2009 *Physica C* **469** 521

[37] Kashiwaya S, Tanaka Y, Koyanagi M, Takashima H and Kajimura K 1995 *Phys. Rev. B* **51** 1350

[38] Krawiec M, Gyöffry B L and Annett J F 2004 *Phys. Rev. B* **70** 134519

[39] Wolf E L 1985 *Principle of electron tunnelling spectroscopy*, Oxford University Press

[40] Pannetier B and Courtois H 2000 *Journ. of Low Temp. Phys.* **118** 599

[41] Nguyen C, Kroemer H and Hu E L 1992 *Phys. Rev. Lett.* **69** 2847

[42] Strijkers G J, Ji Y, Yang F Y, Chien C L and Byers J M 2001 *Phys. Rev. B* **63** 104510